\RequirePackage{fix-cm}
\documentclass{svjour3}                     
\smartqed  
\usepackage{graphicx}
\journalname{}
\begin{document}

\title{The 3D entangled structure of the proton
\thanks{Invited talk presented at the Lightcone 2017 Workshop, 18-22 Sep 2017, Mumbai, India; preprint Nikhef 2018-002, submitted to Few Body Physics}
}
\subtitle{Transverse degrees of freedom in QCD, momenta, spins and more}


\author{P.J. Mulders 
}


\institute{P.J. Mulders \at
              Nikhef Theory Group and Faculty of Science, VU \\
              Science Park 105, NL-1098 XG Amsterdam, the Netherlands \\ 
              \email{p.j.g.mulders@vu.nl}           
}

\date{Received: 10 Jan 2018 / Accepted: date}

\maketitle

\begin{abstract}
Light-front quantized quark and gluon states (partons) play a dominant role in high energy scattering processes. Initial state hadrons are mixed ensembles of partons, while produced pure partonic states appear as mixed ensembles of hadrons. The transition from collinear hard physics to the 3D structure including partonic transverse momenta is related to confinement which links color and spatial degrees of freedom. We outline ideas on emergent symmetries in the Standard Model and their connection to the 3D structure of hadrons. Wilson loops, including those with light-like Wilson lines such as used in the studies of transverse momentum dependent distribution functions (TMDs) may play a crucial role here, establishing a direct link between transverse spatial degrees of freedom and gluonic degrees of freedom. 
\end{abstract}

\section{Introduction}
\label{intro}

Quantum Chromodynamics (QCD) appears to be a distinct part of the standard model describing the strong interactions decoupled from the electroweak interactions. Interactions in the Standard Model are based on local gauge invariance, for QCD linked to the SU(3) color symmetry. At the level of the asymptotic states the color is invisible and free quarks or gluons are absent in the spectrum. On the other hand the color degree of freedom is clearly visible in the notions of valence quarks and in color factors $N_c$ or $1/N_c$ in electron-positron annihilation or Drell-Yan scattering, respectively. Furthermore, it is visible in the factorization of high energy scattering processes in distribution functions $f_{H\rightarrow q}$ and fragmentation functions $D(q\rightarrow h)$ interpretable as momentum densities of partons $q$ in hadrons $H$ and produced number of hadrons $h$ from parton $i$, where the interactions of parton(s) $i$ are described within the Standard Model. Besides these global color features, even the flow of color in the hard process is visible through the appearance of future and past-pointing gauge links that appear in the field theoretical operator definitions~\cite{Collins:1981tt,Belitsky:2003nz,Boer:2003cm} of these parton distribution and fragmentation functions. 

This field theoretical framework is powerful and has been very successful for collinear functions $f(x)$ and $D(z)$ that depend on the light-cone momentum fractions $x = p_q^+/P_H^+$ and $z = P_h^-/k_q^-$, where the incoming and/or outgoing hadrons define the light-like directions $n_+$ and $n_-$. The momentum fractions can be connected to scaling variables, such as the identification of the momentum fraction $x$ with the Bjorken scaling variable $x_B$ in deep inelastic scattering. The partons are identified with the good components of quarks and gluons~\cite{Kogut:1969xa}, the projected fields $\frac{1}{2}\gamma^-\gamma^+\psi$ and $g_T^{\alpha\mu}A_\mu^a$, linking the distribution and fragmentation functions in a natural and easy way to the wave functions in front form quantum mechanics~\cite{Dirac:1949cp}. This link persists if one includes transverse momentum dependence, allowing the parton and hadron momenta to be non-collinear, $p_T = p - xP_H \ne 0$. As long as the kinematics of the hard process is such that the dependence on one of the light-like components of the partons is power suppressed, there are possibilities to separate the hadronic (soft) physics from the partonic (hard) physics. Typically, observables linked to transverse degrees of freedom show up as azimuthal dependences in a non-collinear situation, for instance measured with respect to a transverse polarization direction of the target, requiring polarimetry in the final state or looking at deviations from expected back-to-back situations in production of isolated particles (e.g.\ photons) or the production of jets.


Here, I want to mostly focus on a different view on transverse degrees of freedom involving aspects of entanglement and taking a radical starting point with less dimensions, namely a 1 + 1 dimensional starting point. Although this would in principle have drastic consequences for QCD, it also may provide opportunities of solving long-standing puzzles.
At the end, I come back to transverse degrees of freedom in QCD.

\section{A different view}
\label{sec:1}

In quantum mechanics {\em multipartite} states living in ${\cal H}\otimes{\cal H}\otimes \ldots = {\cal H}^{\otimes N}$ play an important role in entanglement phenomena. For instance bipartite states are known from the EPR experiment. Entanglement is in that case limited to a single class of (Bell) states. For our purpose, it is sufficient that classes are defined via an equivalence under local unitary (LU) transformations, where 'local' refers to an individual Hilbert space that belongs to the direct product multipartite space. When the multipartite space is a multi-particle space such as in the standard EPR discussion, the locality would refer to different particles, but for multipartite states locality can also be a true space-time locality as in the case of the 3D harmonic oscillator considered as the direct product of three 1D harmonic oscillators. For multipartite states there are in general more classes of entangled states. In general an entangled multipartite state leads to mixed ensembles in reduced Hilbert spaces. This might thus apply to hadron structure, when a pure state such as a proton shows up as an ensemble of partonic states (discussed in a slightly different context in Ref.~\cite{Kharzeev:2017qzs}) or the other way around when a pure partonic state fragments into an ensemble of hadrons. We want to go further and actually consider all basic constitutents of the Standard Model, leptons and quarks, as different classes of (entangled) multipartite states. As an aside we note that multipartite entangled states play a role in classical versus quantum behavior with classical states representing particularly suitable bases of entangled multipartite states~\cite{Hooft:2014kka}, possibly governed by rules such as a principle of maximal entanglement~\cite{Cervera-Lierta:2017tdt} which was discussed for a situation of multi-particle entanglement.

We propose specifically to describe the degrees of freedom in the Standard Model as tripartite states and at the same time start with less dimensions. Each local Hilbert space has chiral right (R) and left (L) states, actually represented as right- and left-movers in a 1 + 1 dimensional Minkowski space. Such a reduction of the number of dimensions~\cite{Stojkovic:2014lha} has lots of advantages, e.g.\ no naturalness problem arises because of improved convergence as the scalar fields are dimensionless (in general $(d-2)/2$) and fermion fields have canonical dimension 1/2 (in general $(d-1)/2$). To get a satisfactory description of degrees of freedom and symmetries in the Standard Model we do need the presence of three (real) degrees of freedom.  

To see how space-time dependence and symmetries emerge, we consider the local Hilbert space built with creation and annihilation operators for bosons and fermions satisfying $[a,a^\dagger] = \{b,b^\dagger\} = 1$ with in the case of the presence of additional degrees of freedom a collection of supercharges $Q_{ik} = b_i^\dagger a_k$ and $Q_{ik}^\dagger$ that switch between bosons and fermions. These operators satisfy
\begin{equation}
\{Q_{ik}^\dagger, Q_{jl}\} = \frac{1}{2}\delta_{ij}\{a_l^\dagger,a_k\} + \frac{1}{2}\delta_{kl}[b_i^\dagger,b_j],
\label{maximalsymmetry}
\end{equation}
the right-hand side containing the number operators (trace with $i=j$ and $k=l$) and unitary rotations among the internal degrees of freedom. The number operators can act as the momentum operators or Hamiltonians generating space-time. The creation and annihilation operators are contained in the 	field excitations around the bosonic vacuum, $\varphi = v_0(\phi-1) = (a + a^\dagger)/\sqrt{2\omega}$ and $\xi = (b + b^\dagger)/\sqrt{2}$. With the anti-hermitean supercharge combination $Q = \sqrt{\omega}(b^\dagger a - b a^\dagger)$ one has 
\begin{eqnarray}
&&
\{Q,Q^\dagger\} = \omega(\{a^\dagger,a\} + [b^\dagger,b]) = 2(H+Y),
\label{QQ}
\end{eqnarray}
which contains on the RHS the Hamiltonian (trace) and possible remaing rotations among internal degrees of freedom, schematically indicated as $Y$. For the boson field(s) $\phi$ and fermion field(s) $\xi$ we have 
\begin{eqnarray}
&&
[Q,\phi] = \xi,
\\
&& F = \{Q,\xi\} = \{Q,[Q,\phi]\} = \frac{1}{2}[\{Q,Q\},\phi] = [\phi,H+Y] = iD\phi,
\label{basicDB}
\\
&& [Q,F] = [Q,\{Q,\xi\}] = \frac{1}{2}[\{Q,Q\},\xi] = iD\xi.
\label{basicDF}
\end{eqnarray}
For a single component real field, just the Hamiltonian appears on the RHS of $\{Q,Q^\dagger\} = 2H$ and $[\phi,H] = i\partial_0\phi$. With internal degrees of freedom in the Hilbert space, the possibility of (unitary) transformations is incorporated through the covariant derivative $iD = i\partial + gA$ where $gA$ links locality and the internal transformations among the fields, the basic property of a gauge field. The space-time dependence enters via the Wilson line,
\begin{equation}
\phi(x) = {\cal P} \exp\left(-i\int_0^x ds{\cdot}D(s)\right)\phi.
\label{basic-x}
\end{equation} 
Details, including nontrivial results involving (non-abelian) Wilson loops, still depend on dynamics incorporated in a Lagrangian constrained by symmetry requirements and vacuum structure. The symmetry in the tripartite Hilbert space ${\cal H}^{\otimes 3} = {\cal H}\otimes{\cal H}\otimes {\cal H}$ with imposed Z(3) symmetry among the spaces and in principle three real degrees of freedom is U(3)$_R\times$U(3)$_L$ where the generators of the U(1) parts act as hamiltonians $P^\pm$ that are coupled through the Casimir operator $M^2 = P^+P^-$ or equivalently one has a P(1,1)$\otimes$SU(3) symmetry, where the Poincar\'e symmetry P(1,1) includes $C$ and $P$ and the left and right states belong to conjugate representations that are built on a R-L symmetric vacuum. The choice of light-like components can be local, governed by the choice of a local zweibein $n_\pm(x)$ in Minkowski space. To achieve a consistent picture one also needs the Poincar\'e invariance in field space to avoid problems with the Coleman-Mandula theorem~\cite{Coleman:1967ad}. 
For right and left fields the Wess-Zumino lagrangian~\cite{Wess:1973kz} restricted to 1 + 1 dimension is a nice starting point, being quite general and incorporating supersymmetry as well as the boost-like invariance of the scalar and fermion fields. We extend this to 1 + 3 dimensions~\cite{Mulders:2016ofb,Mulders:2016qve} by extracting the real (space) rotations from the internal symmetry for tripartite states.

\section{Emergent Symmetries of the Standard Model}
\label{sec:2}

\begin{figure}
\includegraphics[width=0.5\textwidth]{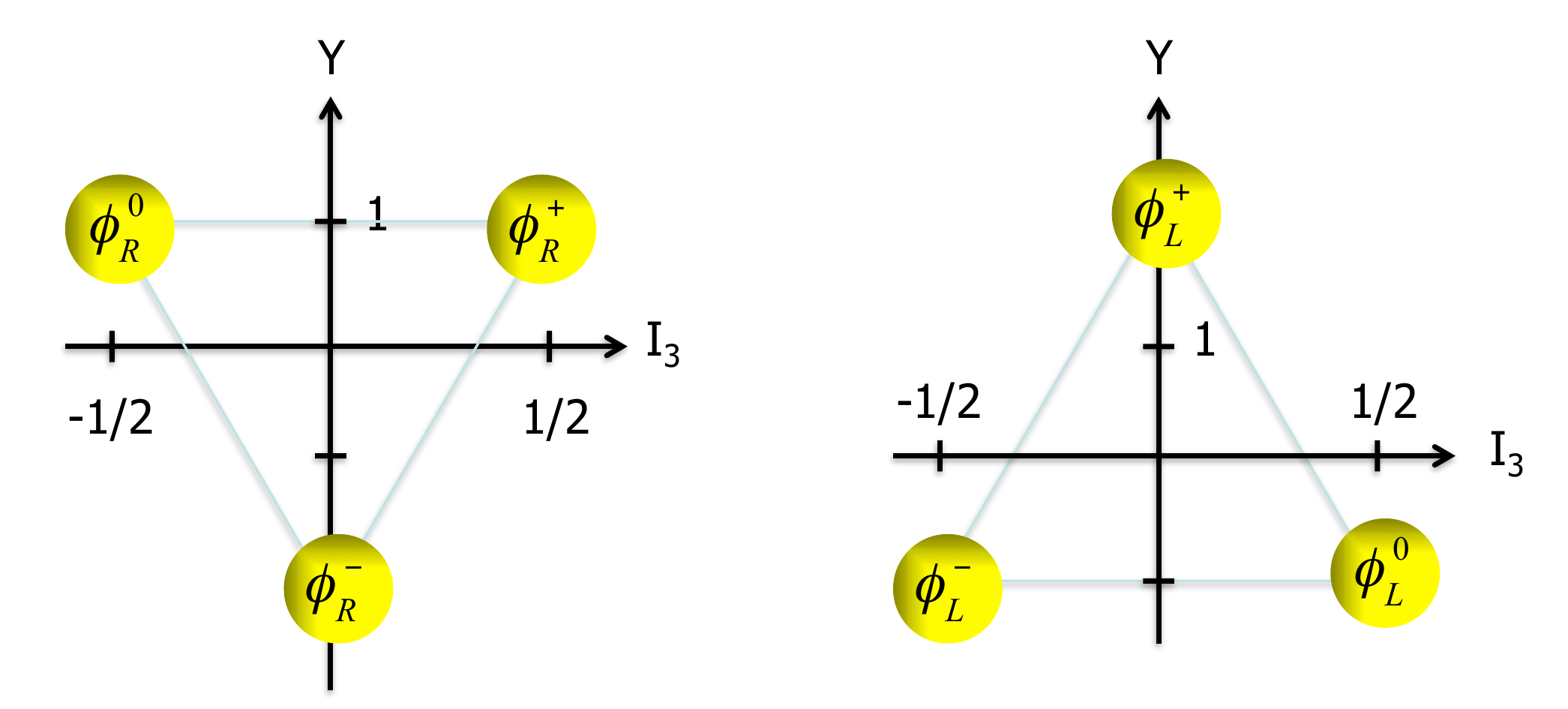}
\includegraphics[width=0.5\textwidth]{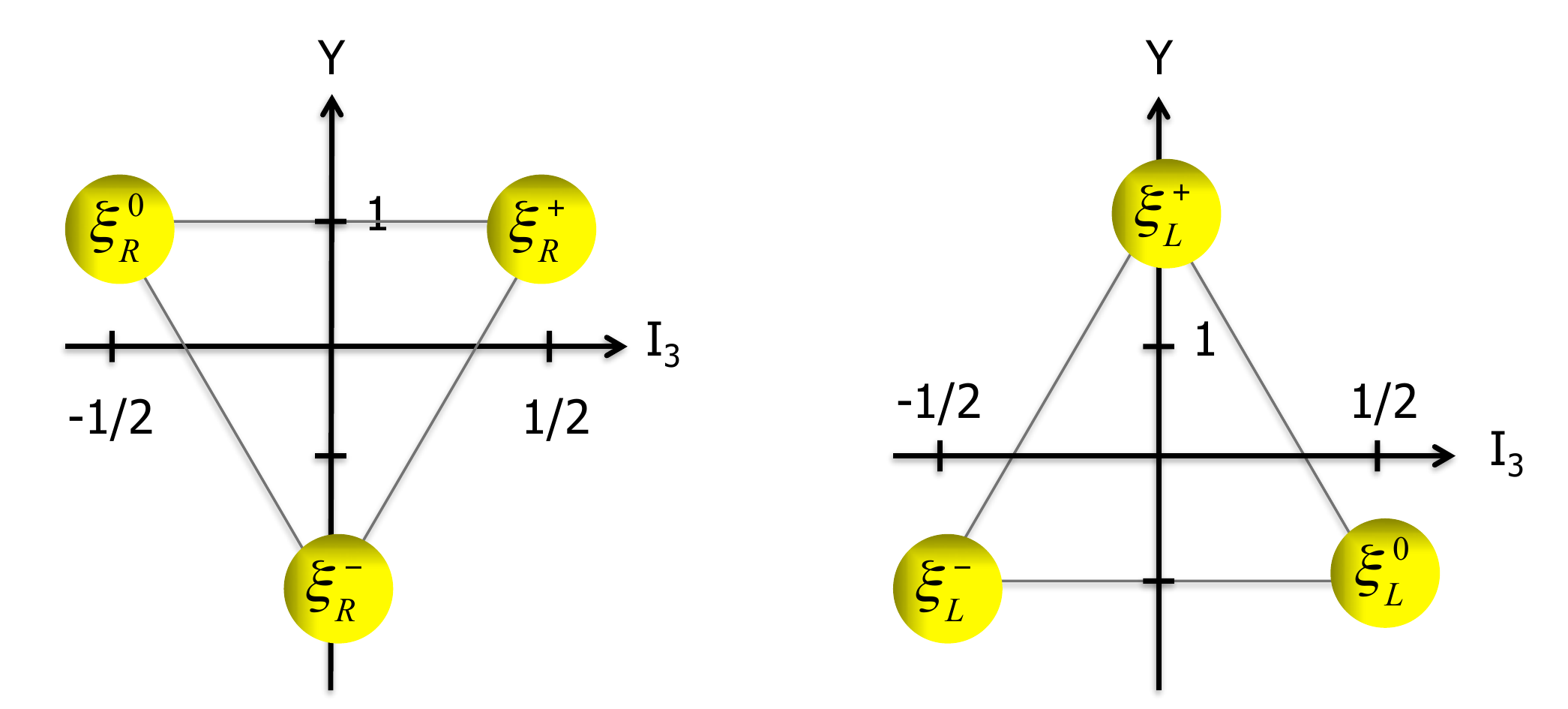}
\caption{Basic triplet/antitriplet of right-handed ($\phi^i = \phi_R^i$) and left-handed bosons ($\bar\phi^i = \phi_L^i$) and the corresponding fermion multiplets. 
Choosing electroweak axes ($Y$, $I_3$), charged and neutral states are identified,
e.g.\ $\phi_R^+$, $\phi_R^0$, $\phi_R^-$.}
\label{basic-ch}       
\end{figure}

In addition to being R or L states, the basic states belong to the triplet representations of SU(3), see Fig.~\ref{basic-ch}. The right and left boson fields can be recoupled to scalar and pseudoscalar fields, contained in a single (complex) field
\begin{eqnarray}
\phi\sqrt{2} &=& e^{i\pi/4}\phi_R + e^{-i\pi/4}\phi_L = \phi_S + i\phi_P,
\label{Lfield-complex}
\end{eqnarray}
(thus $\phi_{R/L} = \phi_S\pm\phi_P$), with nonvanishing expectation value $\langle\phi_S\rangle = v_0$ and $\langle\phi\rangle = v_0/\sqrt{2}$. For fermions we have the spinor field
\begin{eqnarray}
\psi &=& \frac{1}{\sqrt{2}}\left[\begin{array}{c}\xi_R \\ -i\xi_L\end{array}\right].
\end{eqnarray}
For the Wess-Zumino lagrangian the vacuum expectation value is related to the mass $M$ that couples right-left and the (fermion-boson) Yukawa coupling $g_0$ through $v_0 = M/2g_0$ and can be taken $v_0 = 1$.

The vacuum expectation value of the boson fields in tripartite space is R-L symmetric and SO(3) invariant among the (real) degrees of freedom, including Z(3) invariance in tripartite space, which is the center symmetry of SO(3). The remaining unitary symmetry is an SU(2)$\times$ U(1) symmetry with the commutator algebra [SU(2)$\times$U(1)\,,\,SO(3)] closing the SU(3) algebra. The SO(3) and SU(2)$\times$U(1) parts correspond to the 'real rotations' and 'phases', respectively. The phases can be used to assign the electroweak quantum numbers to the basic states in Fig.~\ref{basic-ch}. Since the vacuum is real, the SU(2)$\times$U(1) symmetry, identified as the electroweak symmetry, is broken, leaving only an unbroken U(1)$_Q$ symmetry, of which the generator is identified with the charge operator. 
 
According to Goldstone's theorem the structure of the spectrum of physical states is governed by the symmetries of the vacuum. The bosonic vacuum in ${\cal H}$ is constant, $P^\pm\vert 0\rangle = 0$, implying $\vert\langle \phi_R\rangle\vert = \vert\langle\phi_L\rangle\vert = 1/\sqrt{2}$ thus breaking the R-L symmetry to the diagonal symmetry. In tripartite space the vacuum structure is a Z(3) singlet. The SO(3) symmetry is embedded into the asymptotic P(1,3) symmetry, noting that the commutator algebra [P(1,1)\,,\,SO(3)] closes the P(1,3) algebra. To fully implement the Poincar\'e invariance, in particular the four momentum operators, the zweibein $n_\pm(x)$ must be extended to to a (local) vierbein $n_\mu(x)$ in E(1,3), which can include curvature and classical gravity. 
This extension is governed by the discrete A(4) symmetry group with the Z(3) symmetry group among the tripartite subspaces as a subgroup. The permutations are limited to oriented ones, since P(1,1) already includes parity and charge conjugation. This A(4) embedding symmetry will play a natural role in the family structure of fermions.

Summarizing, depending on the excitations to be studied one includes only boosts (1D) or both boosts and rotations (3D) in the space-time behavior of fields. This leads to two options for the covariant derivatives in Eqs~\ref{basicDB} and \ref{basicDF}, 
\begin{eqnarray}
E(1,1):&& 
i D_\sigma\phi^i = i\partial_\sigma\phi^i + g_0\sum_{a\in G} A_\sigma^a (T_a)^i_j\phi^j,
\label{covderL1}
\\
E(1,3):&&
iD_\mu\phi^i = i\partial_\mu\phi^i 
+ g\sum_{a\in G^\prime} A_\mu^a (T_a)^i_j\phi^j,
\label{covderL3}
\end{eqnarray}
where in the first option all unitary transformations in tripartite space remain part of the internal degrees of freedom (thus $\sigma = 0,1$ and $T_a$'s are generators of full threefold unitary symmetry G = SU(3)), while in the second (asymptotic) option the SO(3) (real) part of the unitary transformations are incorporated in the space-time structure (thus $\mu = 0,1,2,3$ and the generators belong to G$^\prime$ = SU(2)$\times$U(1)). In order to avoid any conflicts with the Coleman-Mandula theorem when implementing the symmetries in Eqs~\ref{basicDB} and \ref{basicDF} to obtain \ref{covderL1} and \ref{covderL3}, it is essential that Poincar\'e invariance is incorporated in the fields, in particular ensuring the appropriate rotational and boost behavior of scalar and vector bosons as well as this behavior for fermions, for instance in E(1,1) the vector boson is a pseudoscalar boson, while it is a vector field in E(1,3).

There are several indications that the emergent picture as sketched in this paragraph is a natural zeroth order scenario for the standard model. The fact that the electroweak symmetry originates from an SU(3) symmetry implies a (zeroth order) weak mixing angle $\sin\theta_W = 1/2$~\cite{Weinberg:1971nd}. The mass relations between vector bosons are as in the standard model. The gauge invariance and spontaneous symmetry breaking leads to the survival of just a scalar Higgs boson after rotating and gauging away the other components from the two (conjugate) Higgs E(1,1) triplets. In the intermediate step one gets the familiar two E(1,3) doublets. A zeroth order relation that emerges from the supersymmetric starting point is $M_Z : M_H : M_{\rm top} = 1/\sqrt{2} : 1 : \sqrt{2}$. With the E(1,1) fermion-boson coupling fixed at $g_0 = M/2$, the normalization of generators in going from Eq.~\ref{covderL1} to \ref{covderL3} suggests in E(1,3) a dimensionless coupling $g^2 = 3/8$ or $\alpha = 3/128\pi\approx 1/134$. 

Also for fermions, the symmetries of the Standard Model emerge naturally for tripartite states with triplets of basic modes. Now the existence of different classes of LU equivalent tripartite states is important. In contrast to bipartite states, there are two classes of tripartites. For two basis states $R$ and $L$, these are represented by the GHZ state and W-state~\cite{GHZ:1999,Bouwmeester:1998iz,Dur:2000zz,Walter:2015},
\begin{eqnarray}
&&
\vert{{\rm GHZ}}\rangle = (\vert{RRR}\rangle + \vert{LLL}\rangle)/\sqrt{2},
\label{GHZ}
\\&&
\vert{{\rm W}}\rangle = (\vert{LRR}\rangle + \vert{RLR}\rangle + \vert{RRL}\rangle)/\sqrt{3}.
\label{W}
\end{eqnarray}
The first state is maximally entangled but fragile and easily gets unentangled by a local operation/measurement, the second is more robust, in general remaining entangled, where we already explained in the Introduction the meaning of 'equivalent' and 'local' in our context.

\begin{figure}
\includegraphics[width=0.5\textwidth]{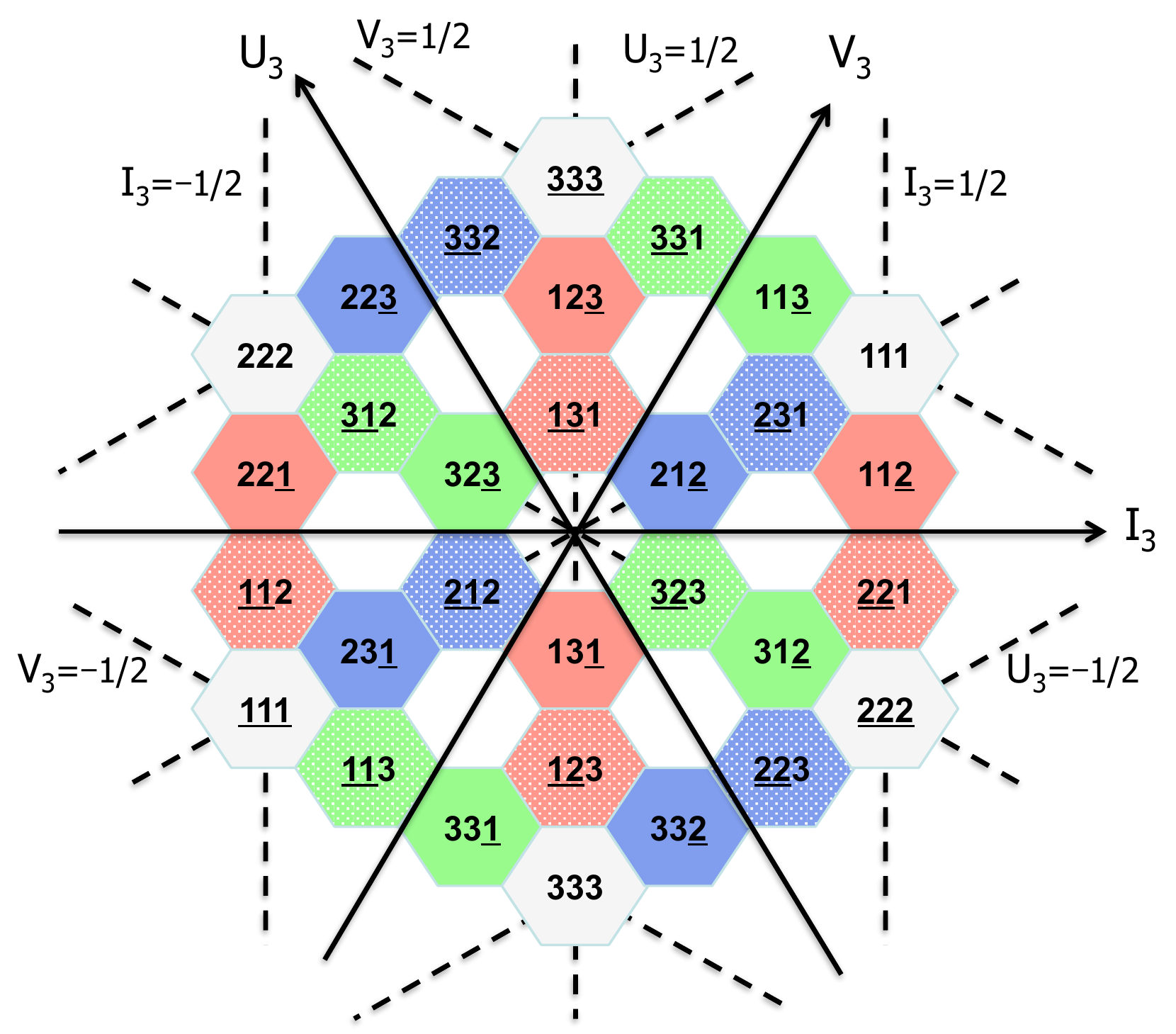}
\includegraphics[width=0.5\textwidth]{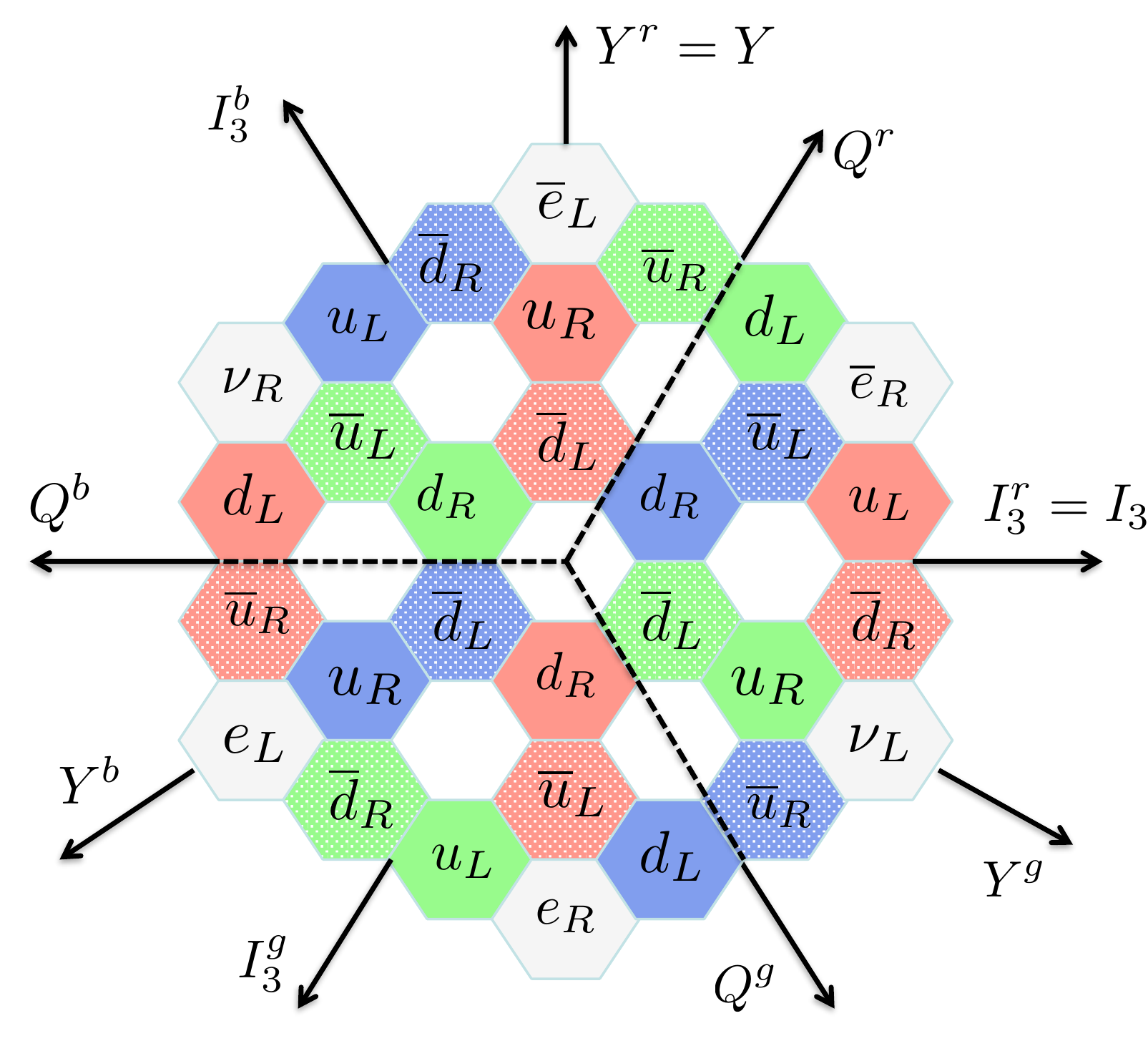}
\caption{Roots for $SU(3)$ that are allowed for GHZ- and W-type tripartite states made up of right (1,2,3) or left (\underline 1, \underline 2, \underline 3) states States with integer or half-integer values of $I$-, $U$- or $V$-spin are shown in the left figure. In the right figure
the asymptotic electroweak assignments of leptons are given by mapping 123 to charged states as in Fig.~\ref{basic-ch}. The assignments of quarks are obtained by freezing color (taking $r$ as reference). }
\label{roots}       
\end{figure}

For the fermionic excitations there are two inequivalent sets of excitations that make use of the full P(1,1) x SU(3) symmetry. The first are {\em asymptotic (3D) states} with electroweak quantum numbers (leptons). The states are aligned (GHZ-type) tripartite states. They belong to one of the three singlet representations of the A(4) embedding symmetry. The three singlet possibilities are identified as the three families, one massive and the others massless. Mass and electroweak eigenstates in zeroth order have a tribimaximal mixing. The electroweak quantum numbers of the aligned states are just the same as those of the fermions in Fig.~\ref{basic-ch}. 

A second set of excitations are {\em non-asymptotic (1D) states} in the Hilbert space of the W-type. The tripartite space is identified as color space. These states are confined quarks and antiquarks. These colored states can come in three different A(4) representations, but now any of the (three) triplet representations, again identified as families. For composite color singlets the rotational symmetry is restored and they can be symptotic 3D states, offering a different view into confinement. For hadrons the local SU(3) symmetry exhibits itself as rotational invariance, a global SU(3) color symmetry, and a local electroweak symmetry. This is the quark valence picture. 

In 1D one has eight gauge fields coupling to the fermions, but there are no dynamical gluons. In fact, just the scalar field coupled to this 1D QCD remains, resembling XQCD in Ref.~\cite{Kaplan:2013dca}. The scalar field in combination with the instantaneous confining (linear) gauge potentials, may play a role in the transition to asymptotic composite colorless states.  In Fig.~\ref{roots} all (aligned and mingled) tripartite fermionic states are given, restricted to allowed states in SU(3) root-space. The six lepton states (aligned states) have allowed I-, U- {\em and} V-spin values, while the 24 quark states (mingled states) have allowed I-, U- {\em or} V-spin values. Since for the 1D quark states I-U-V permutations map onto color permutations, the 24 states constitute 8 triplets/antitriplets of colored quark states. Dynamical gluons appear upon extension of Wilson loops $W[C]$ from 1D into 3D. Interesting in this respect is the link between gluon distribution functions and Wilson loops if one goes beyond collinear QCD and includes transverse momentum dependence~\cite{Boer:2016xqr}.

In order to identify the electroweak quantum numbers of the colored quarks (their {\em valence} nature), the specific identifications $I_3^r = I_3$, $I_3^b = U_3$ and $I_3^g = -V_3$ assure that the electroweak isospin is integer or half-integer. This is illustrated in Fig~\ref{roots}. The resulting allowed SU(2)$\times$U(1) quantum numbers are for each family a left-handed quark doublet and right-handed antiquark doublet and two singlets of opposite handedness. 
The way in which the electroweak structure emerges resembles the rishon model~\cite{Harari:1979gi,Shupe:1979fv,Harari:1981uh}, but rather than having two fractionally charged preons ($V$ and $T$) in $d = 4$, our basic modes are charged or neutral basic 1D excitations. It evades the necessity of compositeness and has some similarities to the effort outlined in Ref.~\cite{Zenczykowski:2008xt}. At a dynamic level, the embedding possibly could connect to a symmetric variable in coordinate space, such as the symmetric light-front variable $\zeta$ in Ref.~\cite{deTeramond:2008ht} using the AdS/CFT correspondence.

\section{Concluding remarks}

We have discussed a different view on the emergence of symmetries and asymptotic degrees of freedom in the Standard Model, even if there are still many open ends. The idea involves a duality between space as part of the Minkowski space and color as the basic tripartite Hilbert space. In the asymptotic world ranging from hadronic to macroscopic and cosmological distances, the excitations are aligned tripartite states. The underlying 1D world emerges at the hadronic scale. We have not yet addressed the emergence of this scale. The picture does away with the confinement issue, quarks are simply not asymptotic states, even if they offer a perfect basis in the appropriate 1D Hilbert space. The 'different view' does not appear to invalidate standard model field theoretical results, although it might affect the way combined (higher order) electroweak and strong corrections are implemented. Depending on how one organizes the Hilbert spaces, it may have links to trinification models (compare e.g.\ \cite{Camargo-Molina:2016bwm}). Within QCD it could provide new ways to look at phenomena like confinement, Bloom-Gilman duality, separation of hard and soft modes in Soft Collinear Effective Theory (SCET), jet physics, color-kinematic duality and the multitude of effective models for QCD including conformal field theory (CFT) approaches. Within the field of parton dynamics, partons being good fields in front form of quantum mechanics, it directly links to the transition of 1D collinear parton distributions to 3D transverse momentum dependent distributions. 

\begin{acknowledgements}
I acknowledge useful discussions with several colleagues at Nikhef and with Fabian Springer who worked on symmetry aspects as part of his MSc project. Except for the presentation at the Lightcone 2017 conference (Mumbai), the work was also presented at the INT 17-03 workshop on Hadron Tomography (INT, Seattle). This research is part of the FP7 EU "Ideas" programme QWORK (Contract 320389) of the European Research Council.
\end{acknowledgements}

\bibliographystyle{spphys}       

%
%

\end{document}